\documentclass[aps,pre,groupedaddress]{revtex4}

\usepackage[dvipdfmx]{graphicx}
\usepackage{mathrsfs}
\usepackage{amsmath,amssymb}
\usepackage{hhline}
\usepackage{dcolumn}
\usepackage{bm}
\usepackage{url}

\usepackage[dvips]{color}

\newcommand{\red}[1]{}

\newcommand{\be}{\begin{equation}}
\newcommand{\ee}{\end{equation}}
\newcommand{\bi}{\begin{itemize}}
\newcommand{\ei}{\end{itemize}}

\newcommand{\ma}{{\rm I}}
\newcommand{\ea}{{\rm D}}

\newcommand{\mn}{m_{\rm N}}
\newcommand{\ms}{m_{\rm S}}

\begin{document}

\title{Cascading Behavior of an Extended Watts Model on Networks}
\date{\today}

\author{Shinji Nishioka}
\affiliation{Graduate School of Science and Engineering, Ibaraki University, 2-1-1, Bunkyo, Mito, 310-8512, Japan}
\author{Takehisa Hasegawa}
\email{takehisa.hasegawa.sci@vc.ibaraki.ac.jp}
\affiliation{Graduate School of Science and Engineering, Ibaraki University, 2-1-1, Bunkyo, Mito, 310-8512, Japan}

\begin{abstract}
In this study, we propose an extended Watts model to examine the effect of initiators on information cascades.
The extended Watts model assumes that nodes with connections to initiators have low adoption thresholds than other nodes, due to the significant influence of initiators.
We develop a tree approximation to describe the active node fraction for the extended Watts model in random networks and derive the cascade condition for a global cascade to occur with a small fraction of initiators.
By analyzing the active node fraction and the cascade window of the extended Watts model on the Erd\H{o}s-R\'{e}nyi random graph, we find that increasing the influence of initiators facilitates the possibility of global cascades, i.e., how many nodes eventually become active is significantly affected by the fraction of initiators and the threshold of nodes directly connected to initiators, which determine cascade dynamics at early stages.
\end{abstract}

\maketitle

\section{Introduction}

Small changes in social networks trigger the spread of opinions, cultural trends, behavioral patterns, and the adoption of innovations (new products and technologies)~\cite{watts2002simple,watts2007influentials}. 
Motivated by suggestive social experiments~\cite{centola2010spread,bakshy2012role,muchnik2013social}, several studies have been devoted to clarifying how social contagion processes cause information cascades~\cite{guilbeault2018complex}.
Watts~\cite{watts2002simple} proposed a mathematical model of information cascades in networks, called the Watts model. 
In the Watts model, individuals on a network can have only two possible states: active or inactive. 
Active means the individual has adopted some innovation; inactive means the individual has not yet adopted the innovation.
Individuals are considered to change their attitude only through social influence---an individual will adopt an innovation if the fraction of its active neighbors exceeds a certain adoption threshold.
As Watts~\cite{watts2002simple} clarified using his model, the adoption of innovation starting from a single initiator ends up with no diffusion, or it spreads across the entire network and becomes a {\it global cascade}, in which a nonzero fraction of the network adopted the innovation, depending on the adoption threshold.
Watts also proposed a condition, in which a global cascade occurs, through the calculation of the percolating vulnerable cluster of a network.

After Watts' seminal work~\cite{watts2002simple}, Gleeson and Cahalane~\cite{gleeson2007seed} proposed a tree approximation for the Watts model on (degree-)uncorrelated networks, which accurately describes cascade dynamics starting from a finite fraction of initiators. 
Based on the tree approximation, they also derived a more accurate cascade condition for triggering a global cascade (extended cascade condition)~\cite{gleeson2007seed}. 
It has also been discussed how the structure of a network affects the probability of triggering a global cascade.
For example, some studies have shown that a global cascade can occur more easily when a network is highly clustered~\cite{ikeda2010cascade,hackett2011cascades,hackett2013cascades} or is positively degree-correlated~\cite{gleeson2008cascades,payne2009information,dodds2009analysis,payne2011exact} than when it is not.
Other studies have investigated the Watts model on modular networks~\cite{galstyan2007cascading,gleeson2008cascades,nematzadeh2014optimal,curato2016optimal}, temporal networks~\cite{karimi2013threshold,takaguchi2013bursty,backlund2014effects}, and multiplex networks \cite{yaugan2012analysis,zhuang2017clustering,unicomb2019reentrant}.

The Watts model has been extended to model complex social contagion processes.
Dodds and Watts \cite{dodds2004universal,dodds2005generalized} proposed a generalized threshold model for biological and social contagion.
Melnik et al \cite{melnik2013multi} investigated a multi-stage contagion model consisting of three states: inactive, active, and hyperactive (hyperactive nodes have a high level of influence compared to active nodes).
In Refs.~\cite{ruan2015kinetics,karsai2016local}, a fraction of nodes is blocked so that they never become active, while other nodes can be activated not only by social influence but also spontaneously become active at random with time. 
Chung et al \cite{chung2019susceptible} classified nodes into three types (influential, susceptible, and normal), with each type having a different influence and adoption threshold: influential nodes have a nonlinear term in their influence, in addition to social passive influence.
Huang et al \cite{huang2016contagion} investigated the Watts model with a persuasion threshold in addition to the adoption threshold to show that a global cascade can occur more easily as the influence of the persuasion threshold is higher.
Other studies have investigated the Watts model with random link weights~\cite{hurd2013watts}, trend followers~\cite{kobayashi2015trend}, degree-dependent thresholds~\cite{lee2017social}, and timers~\cite{oh2018complex}.

Most previous studies have assumed that the adoption threshold of each node is either constant or given from a predetermined probability distribution.
Even in the aforementioned studies~\cite{ruan2015kinetics,karsai2016local,chung2019susceptible} that consider several node types, the type of each node is determined at random.
This assumption also implies that the social influence that an individual receives does not distinguish whether it comes from initiators or others.
Our concern here is {\it how the initiators' influence works in information cascades}.
Initiators will be special and have a higher level of influence than others in that initiators, who took the initiative to adopt an innovation due to its novelty, have the advantage of communicating the benefits of the innovation to others who are entirely unfamiliar with it.
It is then assumed that individuals with connections to initiators have low adoption thresholds than others (who are located farther from initiators).
Little study has been conducted on the impact of initiators' influence on information cascades, even though cascades are caused by a small fraction of initiators.

In this study, we propose an extended Watts model to incorporate a special influence of initiators.
We develop a tree approximation by Gleeson and Cahalane \cite{gleeson2007seed} to describe the active node fraction for the extended Watts model on random networks with arbitrary degree distribution. 
Furthermore, we derive a cascade condition that allows a global cascade to occur in cascades with a small fraction of initiators.
Applying our analysis to the Erd\H{o}s-R\'{e}nyi random graph, we find that increasing the influence of initiators on direct neighbors facilitates the likelihood of global cascades and expands the global cascade region.

\section{Model}

 \begin{figure}[tb]
 \centering
  \includegraphics[width=9cm]{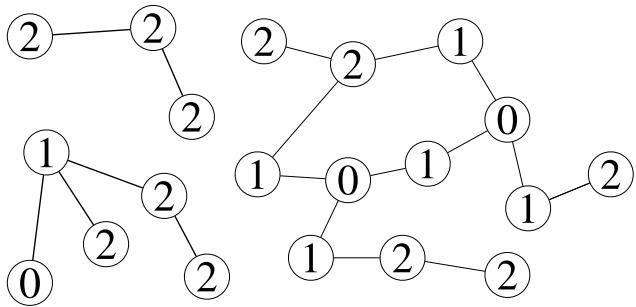}
 \caption{Nodes labeled 0, 1, and 2 represent seeds (initiators), direct neighbors, and indirect neighbors, respectively.
Seeds are randomly chosen from a network in an initial state.
Each non-seed node is a direct neighbor if it is connected to one or more seeds, and is an indirect neighbor if it is not connected to any seed.
}
 \label{fig1}
 \end{figure}

We begin by introducing the (original) Watts model \cite{watts2002simple}.
Let us consider a network with degree distribution $p(k)$.
For each node $i$, we assign the adoption threshold $\theta_i$ according to a predetermined distribution $f(\theta_i)$.
Each node is in one of two states: active or inactive.
In the initial state, a small fraction of nodes $\rho_0(\ll1)$ is randomly chosen to be active and the other inactive.
These initially activated nodes are called initiators or seeds.
At each step, the state of the inactive node $i$ of degree $k_i$ is changed to active if the number, $m_i$, of its active neighbors satisfies $m_i/k_i \geq \theta_i$, where $\theta_i$ is the adoption threshold of node $i$.
The adoption threshold quantifies the sensitivity of a node to the social influence. 
Fewer active neighbors are required for a node to be active when its threshold is smaller.
Once a node becomes active, it is never deactivated.
The process starting from an initial state progresses to a final state.
The mean fraction of active nodes in the final state is called the active node fraction $\rho_\infty$.
A global cascade is stated to occur when a small fraction of seeds yields a considerably greater activation, such that most nodes are eventually activated. 
The model on a network will show $\rho_\infty \gg \rho_0$ when global cascades can occur; otherwise, it will show $\rho_\infty \simeq  \rho_0$.

In order to analyze the impact of seeds in information cascades, we extend the Watts model as follows.
Given a random seed placement, we refer to a non-seed node as a {\it direct neighbor} if it is connected to one or more seeds, and as an {\it indirect neighbor} if it is not connected to any seeds (Fig.~\ref{fig1}).
We assume that direct and indirect neighbors have different distributions of adoption thresholds: threshold $\theta_i$ for node $i$ is independently given from the distribution $f_\ea (\theta_i)$ if it is a direct neighbor, and given from the distribution $f_\ma(\theta_i)$ if it is an indirect neighbor.
For simplicity, this study assumes that $f_\ea(\theta_i)=\delta(\theta_i-\theta_\ea)$ and $f_\ma(\theta_i)=\delta(\theta_i-\theta_\ma)$, i.e., $\theta_i=\theta_\ea$ for all direct neighbors and $\theta_i=\theta_\ma$ for all indirect neighbors.
We note that the extended Watts model reduces to the original Watts model when $\theta_\ea=\theta_\ma=\theta$ (more generally, $f_\ea(\theta)=f_\ma(\theta)=f(\theta)$).
The extended Watts model allows us to consider the situation where each node takes a different threshold depending on whether or not it is adjacent to seeds.
The model defines that $\theta_\ea$ is smaller than $\theta_\ma$ if the influence of seeds is greater than that of others.
In the following section, we investigate how the influence of seeds can enhance the likelihood of global cascades.

\section{Results}

\subsection{Active Node Fraction}

\begin{figure}[tb]
\centering
\includegraphics[width=5cm]{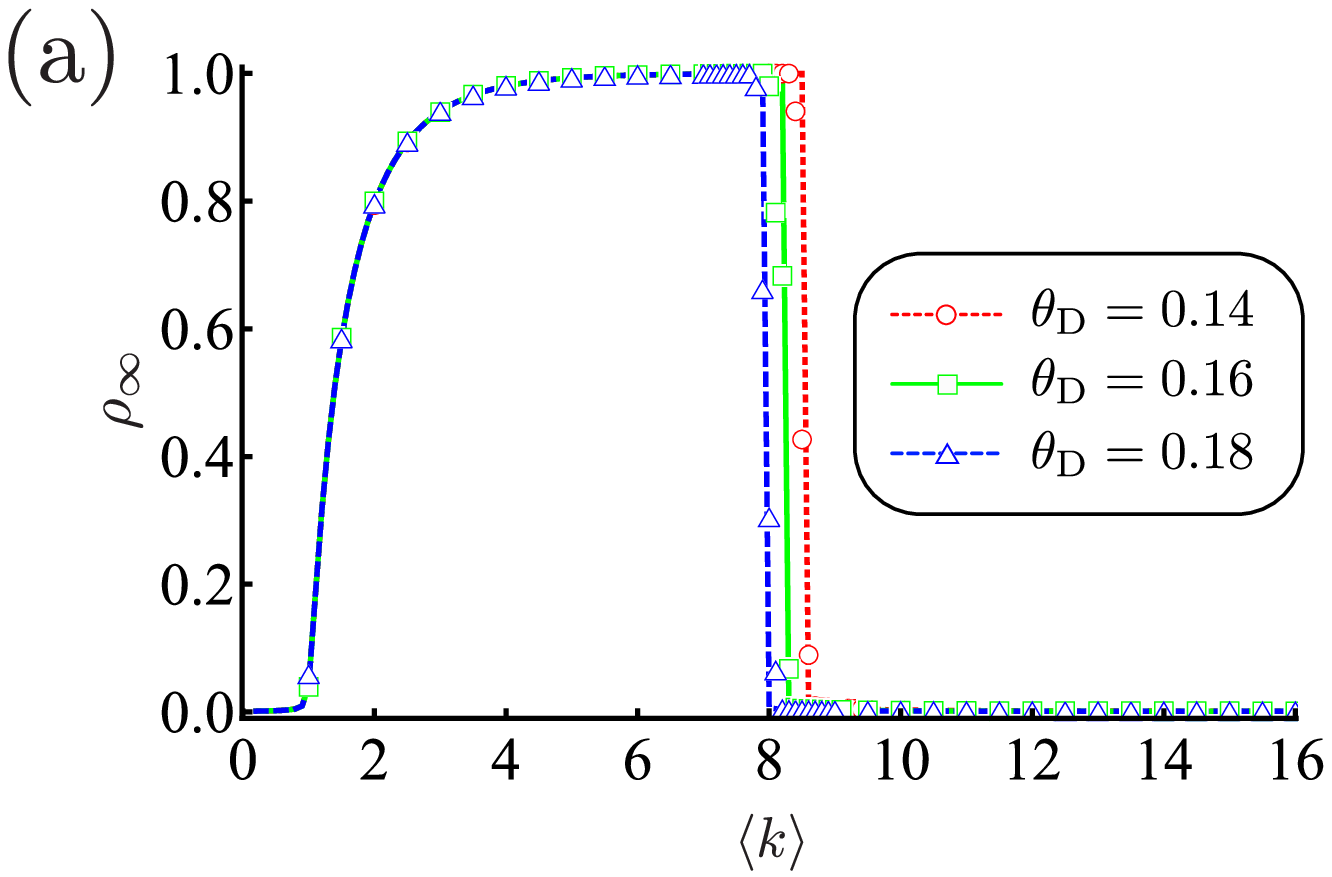}
\includegraphics[width=5cm]{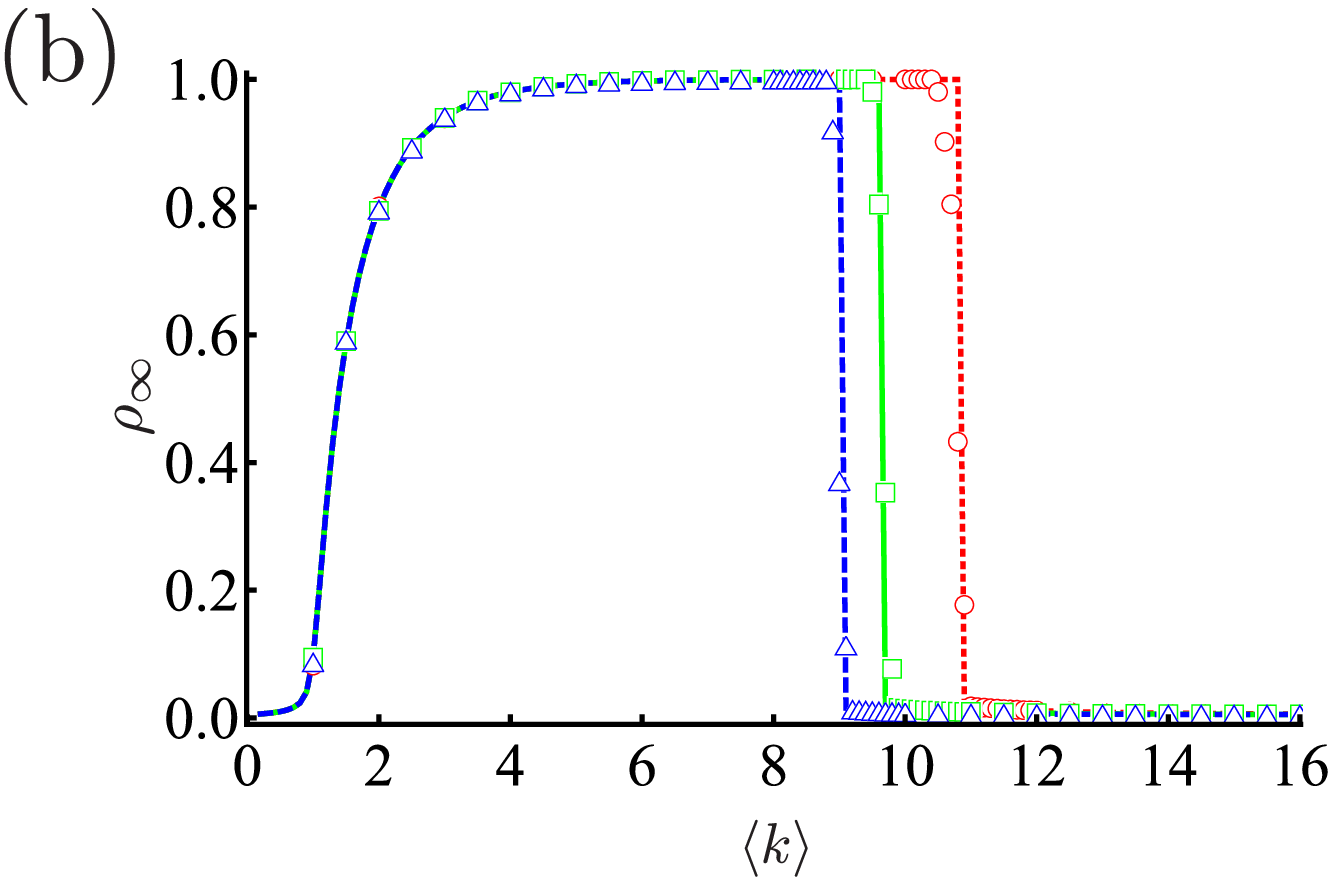}
\includegraphics[width=5cm]{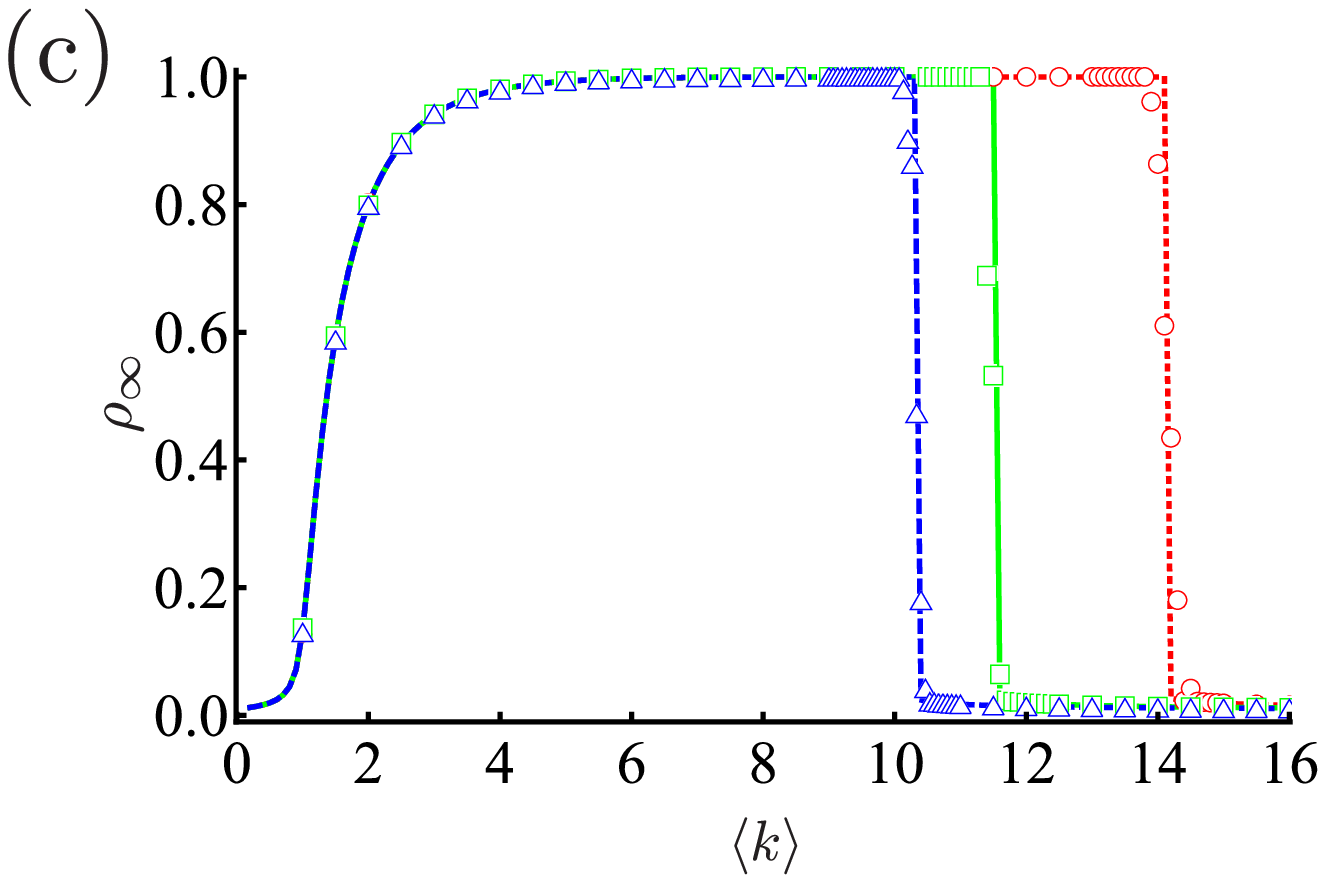}
\caption{
Active node fraction $\rho_\infty$ as a function of $\langle k\rangle$ for the extended Watts model on ERRG under different initial condition: (a) $\rho_0=0.001$, (b) $\rho_0=0.005$, and (c) $\rho_0=0.01$.
In each panel, the threshold for indirect neighbors is $\theta_\ma=0.16$, whereas the threshold for direct neighbors is $\theta_\ea=0.14$ (the red-dotted line and circles), $0.16$ (the green-solid line and squares), and $ 0.18$ (the blue-dashed line and triangles).
The lines represent the analytical results based on the tree approximation (Eq.~(\ref{rho}) and Eq.~(\ref{eq:recEq-rl-alt})), and the symbols represent the Monte Carlo simulation results.
In Monte Carlo simulations, $50$ ERRGs with $N=10^5$ nodes are generated for each parameter, one cascade trial is run for each network, and the active node fraction is averaged over all samples.
}
\label{fig2}
\end{figure}

We develop a tree approximation~\cite{gleeson2007seed} to obtain the active node fraction $\rho_\infty$ for the extended Watts model.
Let us assume an infinitely large (degree-)uncorrelated network with degree distribution $p(k)$.
In an uncorrelated network, the probability that a randomly chosen neighbor of a node has degree $k$ is $kp(k)/\langle k\rangle$.
The active node fraction $\rho_\infty$ defined as the probability that a randomly chosen node is active in the final state is determined as follows:
\begin{eqnarray}
\rho_\infty=&\rho_{0}& \nonumber \\
&+&\left(1-\rho_{0}\right) \sum_{k=1}^{\infty} p(k) \sum_{\ms=1}^{k} \binom{k}{\ms}\, \rho_{0}^{\ms}\left(1-\rho_{0}\right)^{k-\ms} \sum_{\mn=0}^{k-\ms}\binom{k-\ms}{\mn}\, r_{\infty}^{\mn}\left(1-r_{\infty}\right)^{k-\ms-\mn }F_\ea\left(\frac{\ms+\mn}{k}\right) \nonumber \\
&+&\left(1-\rho_{0}\right) \sum_{k=1}^{\infty} p(k)\left(1-\rho_{0}\right)^{k} \sum_{\mn=0}^{k}\binom{k}{\mn}\, r_{\infty}^{\mn}\left(1-r_{\infty}\right)^{k-\mn} F_\ma\left(\frac{\mn}{k}\right), \label{rho}
\end{eqnarray}
where $\ms$ is the number of adjacent seeds, $\mn$ is the number of active adjacent non-seed nodes (i.e., direct and indirect neighbors), $F_{\ea}(x)$ and $F_{\ma}(x)$ are the response functions of direct neighbors and indirect neighbors, respectively, and $r_\infty$ is the probability that a randomly chosen neighbor of a non-seed node is active in the final state.
The response function is as
\be
F_{\rm \ea(\ma)}(x)=\left\{\begin{array}{l}{1~~~~{\rm if} ~~x \ge \theta_{\rm \ea(\ma)}} \\ {0~~~~{\rm otherwise}}\end{array}\right..
\ee
The first term, $\rho_0$, on the right-hand side of Eq.~(\ref{rho}) is the probability that a randomly chosen node is a seed, the second term is the probability that a randomly chosen node is directly connected to seeds and was changed to be active by the active neighbors, and the third term is the probability that a randomly chosen node is not directly connected to seeds and was changed to be active by the active neighbors.

The probability $r_\infty$ that a randomly chosen neighbor of a non-seed node is active in the final state is given as the solution of the recursive equation,
\begin{eqnarray}
r_{l+1}
&=&
\sum_{k=1}^{\infty} \frac{(k+1)p(k+1)}{\langle k\rangle} \sum_{\ms=1}^{k} \binom{k}{\ms}\, \rho_{0}^{\ms}\left(1-\rho_{0}\right)^{k-\ms} \sum_{\mn=0}^{k-\ms} \binom{k-\ms}{\mn}\, r_{l}^{\mn}\left(1-r_{l}\right)^{k-\ms-\mn}F_\ea\left(\frac{\ms+\mn}{k+1}\right) \nonumber \\
&&+
\sum_{k=0}^{\infty} \frac{(k+1)p(k+1)}{\langle k\rangle}\left(1-\rho_{0}\right)^{k} \sum_{\mn=0}^{k} \binom{k}{\mn}\,  r_{l}^{\mn}\left(1-r_{l}\right)^{k-\mn}F_\ma\left(\frac{\mn}{k+1}\right) \nonumber \\ 
&=& 
\sum_{k=0}^{\infty} \frac{(k+1)p(k+1)}{\langle k\rangle} \sum_{m=0}^k \binom{k}{m}\, [(1-\rho_0)(1-r_l)]^{k-m} [\rho_0+(1-\rho_0)r_l]^m F_\ea \left(\frac{m}{k+1}\right) \nonumber \\
&&
+\sum_{k=0}^{\infty} \frac{(k+1)p(k+1)}{\langle k\rangle}\left(1-\rho_{0}\right)^{k} \sum_{m=0}^{k} \binom{k}{m}\, r_l^{m}(1-r_l)^{k-m} \left[F_\ma\left(\frac{m}{k+1}\right) - F_\ea\left(\frac{m}{k+1}\right)\right], \label{eq:recEq-rl-alt}
\end{eqnarray}
where the initial condition is $r_0=0$.
Evaluating $r_\infty$ from iterative computations of Eq.~(\ref{eq:recEq-rl-alt}) and substituting it into Eq.~(\ref{rho}), we obtain the active node fraction $\rho_\infty$ of the extended Watts model with arbitrary parameters ($\rho_0,~\theta_\ea,~\theta_\ma$) on a given network.

To recognize typical behaviors of the extended Watts model, this study uses the Erd\H{o}s-R\'{e}nyi random graph (ERRG), whose degree distribution follows $p(k) = e^{-\langle k \rangle}\langle k\rangle^k/{k!}$.
Figure~\ref{fig2} shows the active node fraction $\rho_\infty$ as a function of the average degree $\langle k \rangle$ of the underlying ERRG.
With fixing the threshold for indirect neighbors at $\theta_\ma=0.16$, three different values of the threshold for direct neighbors are considered: $\theta_\ea=0.14, 0.16, 0.18$.
Furthermore, to validate our tree approximation, we performed Monte Carlo simulations. 
The fact that the theoretical lines for the active node fraction $\rho_\infty$ match well with simulation data (symbols) in all cases shows that the current tree approximation accurately describes the cascade dynamics of the extended Watts model.

When $\langle k \rangle$ increases from zero, we observe transitions in $\rho_\infty$ at a smaller $\langle k \rangle$ and a larger $\langle k \rangle$, respectively.
First, when $\langle k \rangle$ exceeds $\langle k \rangle_{\rm c1}=1$, $\rho_\infty$ increases continuously from $\rho_\infty \approx \rho_0$.
The transition at $\langle k \rangle=\langle k \rangle_{\rm c1}$ is thought of as the phase transition {\it of} the ERRG and is independent of thresholds $\theta_\ea$ and $\theta_\ma$.
For $\langle k \rangle <\langle k \rangle_{\rm c1}$, the ERRG has no macroscopic components, and any cascades are restricted within small components.
In an intermediate range of $\langle k \rangle$, $\langle k \rangle_{\rm c1}< \langle k \rangle<\langle k \rangle_{\rm c2}$, a small fraction of seeds can trigger a global cascade, resulting in $\rho_\infty>\rho_0$.
At $\langle k \rangle=\langle k \rangle_{\rm c2}$, the active node fraction drops discontinuously from $\rho_\infty \approx 1$ to $\rho_\infty \approx \rho_0$.
For $\langle k \rangle>\langle k \rangle_{\rm c2}$, cascades starting from a small number of seeds hardly spread in that each node, on average, has many inactive neighbors (it will not become active even if one or two of its neighbors are active).
The same behavior has been reported for the original Watts model~\cite{watts2002simple, gleeson2007seed}.

The higher critical value $\langle k \rangle_{\rm c2}$ depends on both the seed fraction $\rho_0$ and two thresholds, $\theta_\ea$ and $\theta_\ma$.
As shown in Fig.~\ref{fig2}, when $\rho_0$ and $\theta_\ma$ are fixed, the global cascade region becomes wider ($\langle k \rangle_{\rm c2}$ is higher) as $\theta_\ea$ decreases.
It can be deduced that increasing the influence of seeds on direct neighbors facilitates the likelihood of global cascades.
A comparison of the results of different seed fractions suggests that an increase in the number of initially activated seeds also increases the likelihood of global cascades.
Furthermore, it increases the number of direct neighbors, which can synergistically enhance the effect from the increase of seeds' influence, as evidenced by the widening of the gap between the red-dotted and blue-dashed lines as $\rho_0$ increases.

\subsection{Cascade Condition}

Next, we derive the cascade condition for a global cascade to occur, assuming that the seed fraction $\rho_0$ is small ($\rho_0 \ll1$).
A second-order Taylor expansion of Eq.~(\ref{eq:recEq-rl-alt}) at $r_l\simeq0$ yields (see Appendix):
\be
r_{l+1} \approx C_0 + C_1 r_l + C_2 r_l^2, \label{eq:rnEq}
\ee
where
\begin{eqnarray}
C_n&=&
\sum_{k=n}^{\infty} \frac{(k+1)p(k+1)}{\langle k\rangle}\sum_{i=0}^n\sum_{m=1}^{k-n}\binom{k}{n} \binom{k-n}{m}\binom{n}{i}(-1)^{i+n}\rho_0^{m} \left(1-\rho_{0}\right)^{k-m}F_\ea\left(\frac{m+i}{k+1}\right)  \nonumber \\
&&+\sum_{k=0}^{\infty} \frac{(k+1)p(k+1)}{\langle k\rangle}\sum_{i=0}^n\binom{k}{n}\binom{n}{i}(-1)^{i+n}\left(1-\rho_0\right)^{k}F_\ma \left(\frac{i}{k+1}\right). \label{Cn}
\end{eqnarray}
According to the argument in~\cite{gleeson2007seed}, when Eq.~(\ref{eq:rnEq}) is linearized as
\be
r_{l+1}=C_0+C_1r_{l},
\ee
the global cascade occurs if
\be
C_1-1>0,  \label{C1}
\ee
for which $r_l$ increases monotonically with $l$ and approaches $1$.
Furthermore, considering Eq. (\ref{eq:rnEq}) at $r_{l+1}=r_l=r_\infty$, we obtain
\be
r_{\infty}=C_0+C_1r_{\infty}+C_2r_{\infty}^2.\label{2globalcascade}
\ee
A nontrivial solution $r_\infty>0$ exists, meaning that a global cascade occurs, when Eq. (\ref{2globalcascade}) satisfies the discriminant $D=b^2-4ac<0$ for the quadratic equation $ax^2+bx+c=0$.
This condition is expressed as
\be
\left(C_1-1\right)^2-4C_0C_2<0. \label{C2}
\ee
To summarize, a global cascade occurs when condition (\ref{C1}) or condition (\ref{C2}) is satisfied.
When $\theta_\ea=\theta_\ma$, implying that the influence of seeds is the same as that of non-seed nodes and the model reduces to the original Watts model, our cascade condition corresponds to the generalized cascade condition for the original Watts model, which was recently developed by Kobayashi and Onaga~\cite{kobayashi2022dynamics}.

 \begin{figure}[t] 
 \centering
\includegraphics[width=5cm]{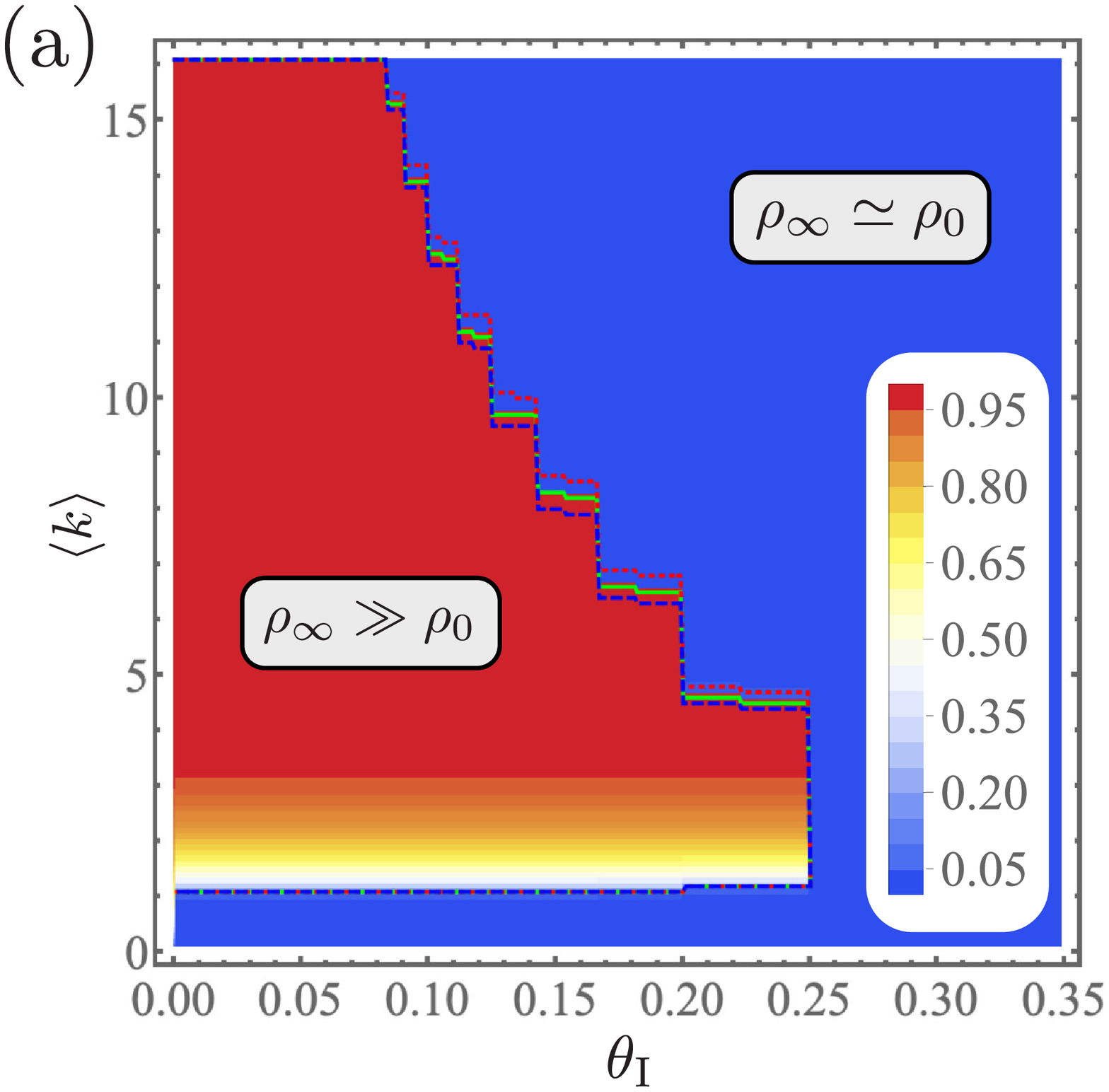}
\includegraphics[width=5cm]{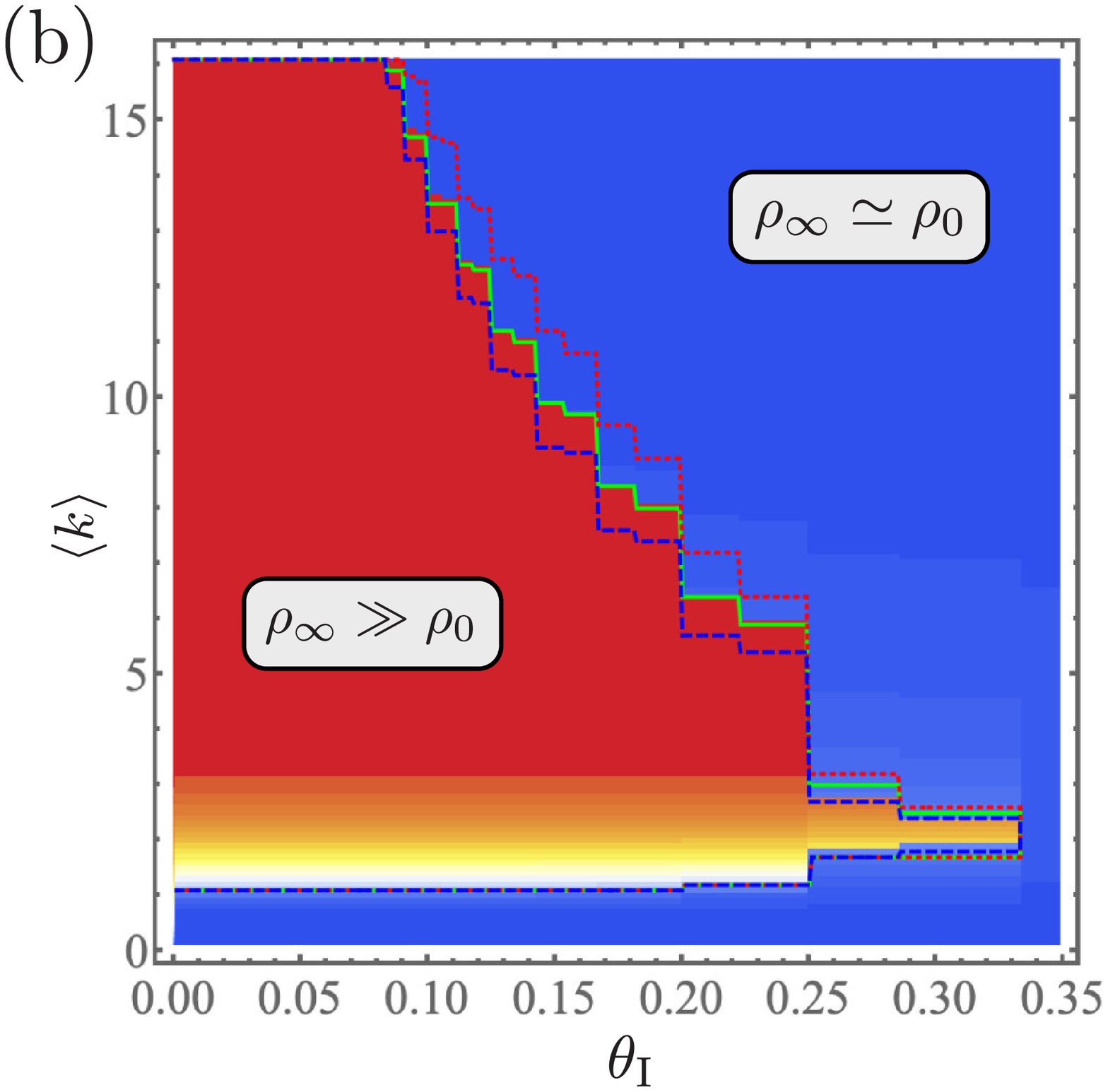}
\includegraphics[width=5cm]{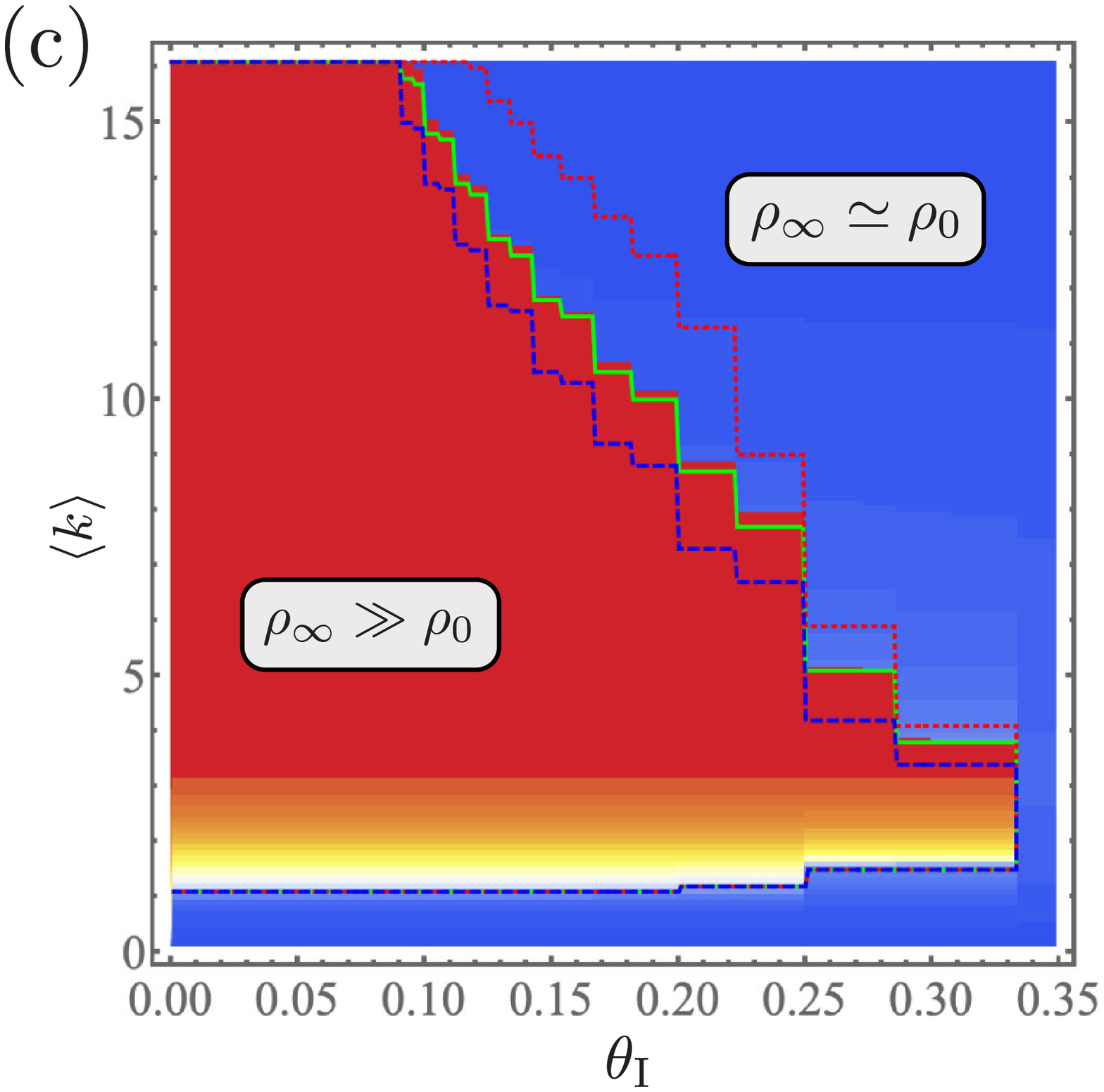}
\caption{
Cascade window in $(\theta_\ma, \langle k \rangle)$ plane for the extended Watts model on the ERRG when the threshold for direct neighbors is $\theta_\ea=0.16$ and the initial conditions are (a) $\rho_0=0.001$, (b) $\rho_0=0.005$, and (c) $\rho_0=0.01$.
Here color-coded values of the active node fraction $\rho_\infty$ are evaluated from Eq.~(\ref{rho}) and Eq.~(\ref{eq:recEq-rl-alt}).
The red-dotted, green-solid, and blue-dashed lines represent the boundary of the global cascade region for cases of $\theta_\ea=0.14$, $0.16$, and $0.18$, respectively.
The cascade condition (Eq.~(\ref{C1}) or Eq.~(\ref{C2})) provides these boundaries.
}
\label{fig3}
\end{figure}

 \begin{figure}[t] 
 \centering
 \includegraphics[width=5.cm]{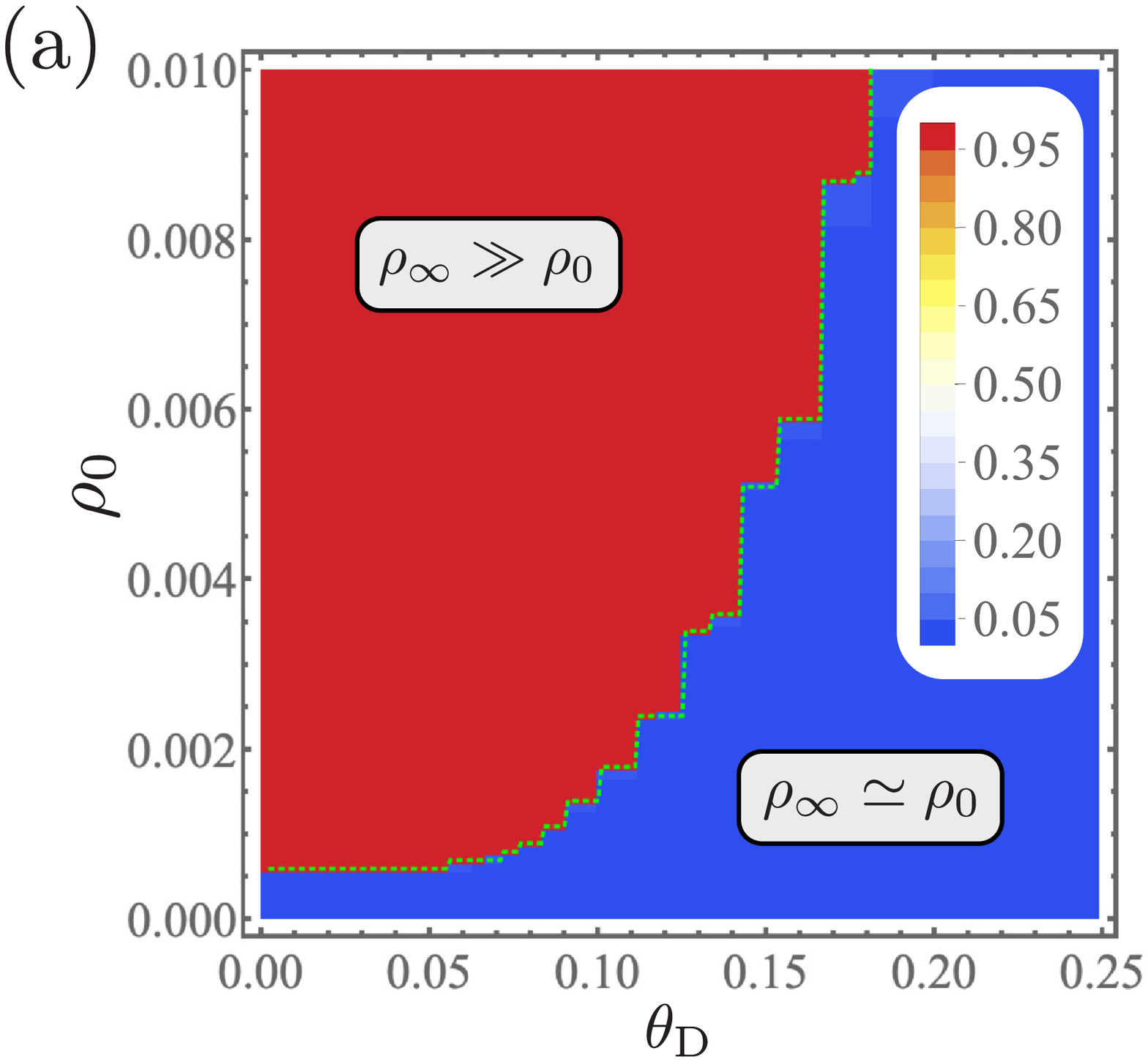}
 \includegraphics[width=5.cm]{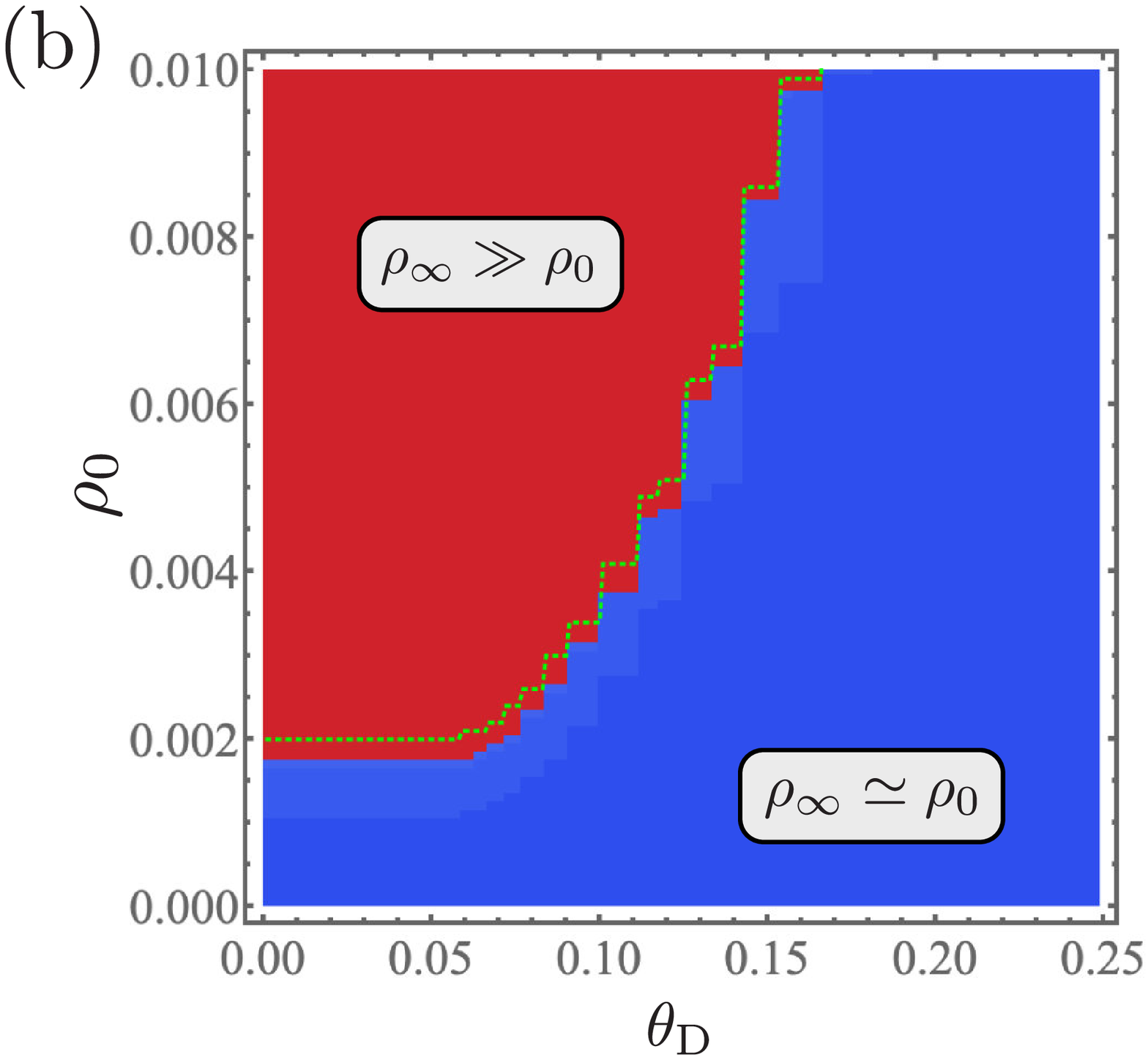}
 \caption{
Impact of seeds' influence on the active node fraction $\rho_\infty$ of the extended Watts model on the ERRG with $\langle k \rangle=10$ and with (a) $\theta_\ma=0.16$ and (b) $\theta_\ma=0.20$. 
Here color-coded values of the active node fraction $\rho_\infty$ are evaluated from Eq.~(\ref{rho}) and Eq.~(\ref{eq:recEq-rl-alt}). 
The global cascade region's boundary (green-dotted line) is given from the cascade condition (Eq.~(\ref{C1}) or Eq.~(\ref{C2})).
}
\label{fig4}
 \end{figure}

Figures~\ref{fig3}(a)--\ref{fig3}(c) illustrate the color-coded values of the active node fraction $\rho_\infty$ with the boundary (the green-solid line) of the global cascade region, which is given by the cascade condition (Eq. (\ref{C1}) or Eq. (\ref{C2})) when the threshold for direct neighbors is fixed at $\theta_\ea=0.16$.
It is clear from the comparison of the colored maps and green-solid lines that our cascade condition accurately describes the boundary between regions where global cascades occur and where they do not.
Furthermore, boundaries for cases of $\theta_\ea=0.14$ (red-dotted lines) and $0.18$ (blue-dashed lines) are also plotted in Figs.~\ref{fig3}(a)--\ref{fig3}(c).
Comparing three boundaries tells us that the global cascade region is expanded as the threshold for direct neighbors becomes lower.
It can also be presumed that as $\langle k \rangle$ and $\rho_0$ become larger, the number of direct neighbors gets larger, leading to an increase in the effect of seeds on the cascade window.
Furthermore, we illustrate the effect of the influence that seeds have on the cascade window.
Figure~\ref{fig4}, showing the cascade window of the $(\theta_\ea, \rho_0)$ plane at fixed $\theta_\ma$ and $\langle k \rangle$, clearly indicates that lowering the threshold for direct neighbors or increasing the seed fraction causes global cascades.
As illustrated by all these cases, the fraction of seeds and the threshold of nodes directly connected to a seed, which mostly determine cascade dynamics at early stages, have a significant impact on how many nodes eventually become active.

\section{Summary}

 \begin{figure}[t] 
 \centering
 \includegraphics[width=5.5cm]{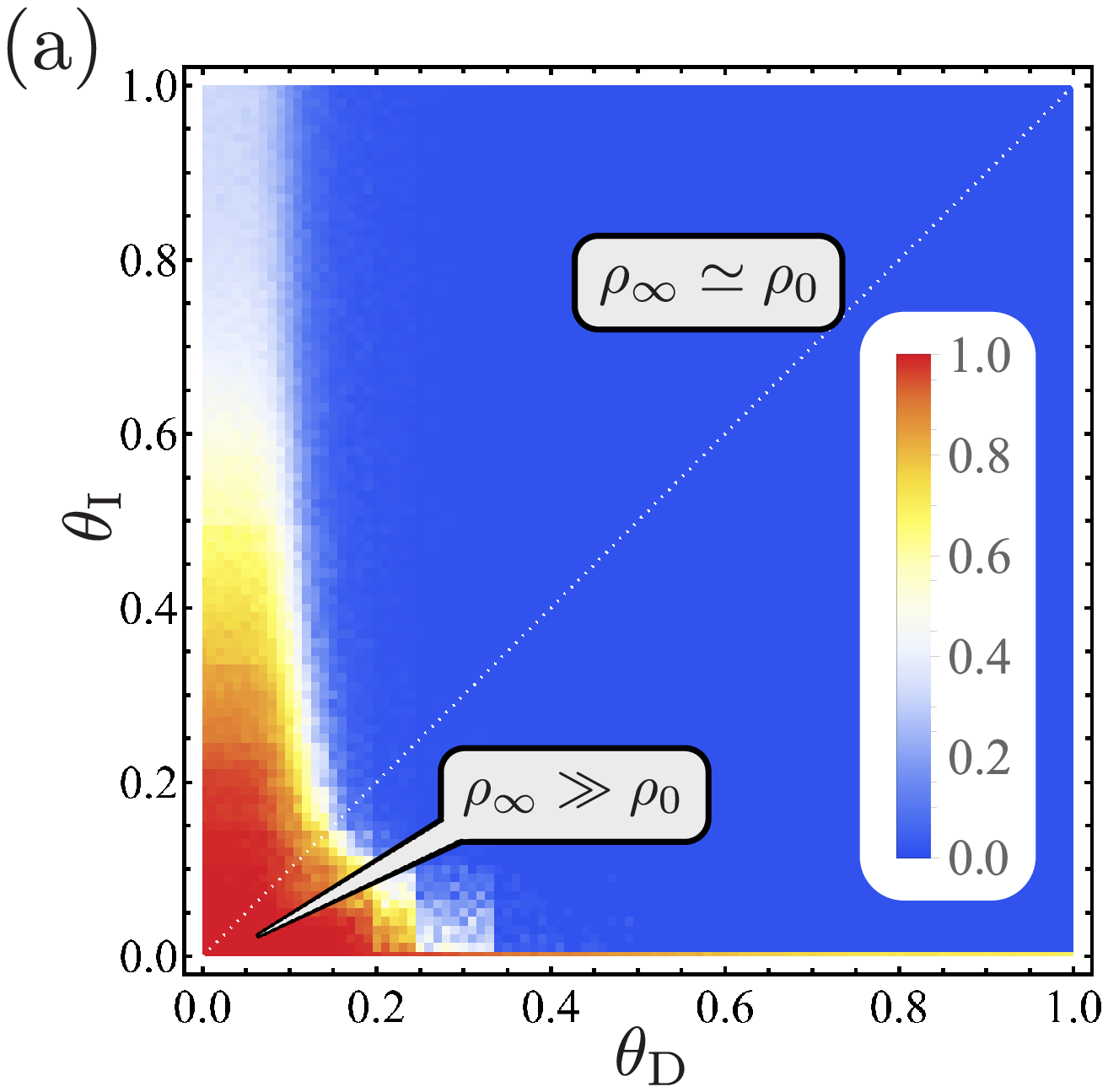}
 \includegraphics[width=5.5cm]{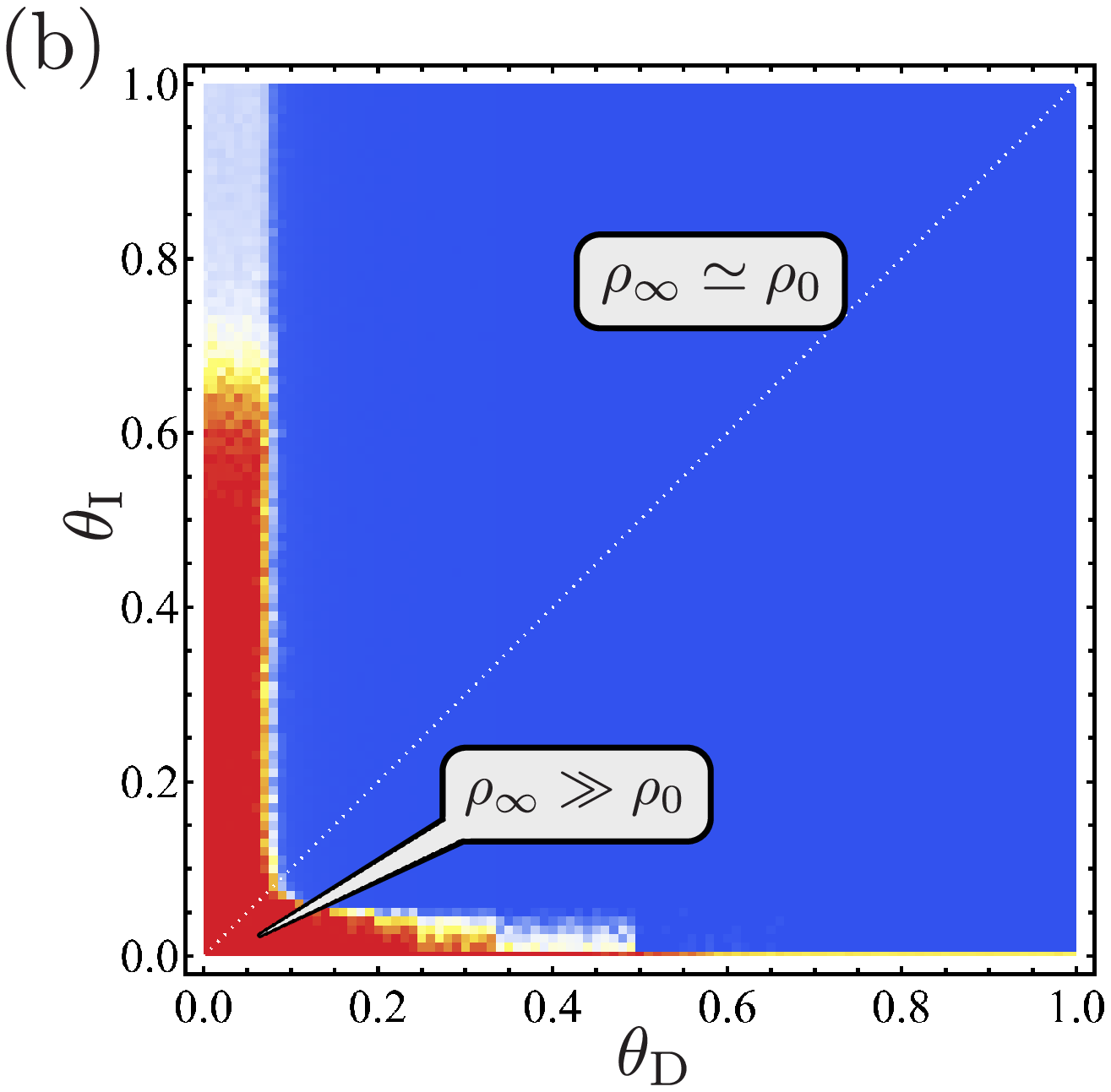}
 \caption{
Active node fraction $\rho_\infty$ of (a) the Facebook network and (b) randomized networks preserving degree of each node in the $(\theta_\ea, \theta_\ma)$ plane.
Both networks have $N=4048$ and $\langle k\rangle\approx43$.
At each combination of $\theta_\ea$ and $\theta_\ma$, $\rho_\infty$ is averaged over 50 samples of Monte Carlo simulations with initial condition $\rho_0=0.01$.
}
\label{fig5}
 \end{figure}

In this study, we proposed an extended Watts model to investigate how the influence that initiators (seeds) have works in information cascades.
The extended Watts model assumes that individuals (direct neighbors) who have connections to initiators have low adoption thresholds than others (indirect neighbors), reflecting the situation that initiators are special and have a higher level of influence than others.
We developed a tree approximation to describe the active node fraction of the extended Watts model on random networks. 
We also derived a cascade condition that allows a global cascade to occur in cascades starting with a small fraction of initiators.
By analyzing the extended Watts model on ERRG, we found that increasing the influence of initiators (lowering the adoption threshold of direct neighbors) facilitates the likelihood of global cascades and expands the global cascade region.

In this study, we concentrated only on the ERRG.
The impact of initiators on cascade dynamics will be obvious in real networks.
Figure~\ref{fig5} shows the active node fraction of an ego network in Facebook~\cite{mcauley2012learning} (Facebook network) and randomized networks preserving degree of each node in the ($\theta_\ea,\;\theta_\ma$) plane.
In both cases, decreasing the threshold for direct neighbors $\theta_\ea$ extends the cascade window to the larger region of the threshold for indirect neighbors $\theta_\ma$.
Furthermore, we see that except for the region $\theta_\ea \gg \theta_\ma$, the global cascade region of the Facebook network is wider than that of the randomized networks when comparing the two cases.
This is because the Facebook network is highly clustered (the clustering coefficient of this network is $C \approx 0.605$ and that of the randomized network is $C \approx 0.068$) and direct neighbors of each node tend to be connected, resulting in that a positive feedback effect on direct neighbors works.
The influence of initiators in the information cascade will be enhanced in other real networks because they are highly clustered.

This study put its focus on a model in which a node's sensitivity to the activation depends on whether or not it is connected to initiators.
To discuss more realistic information cascade dynamics, we may generalize the Watts model so that the adoption threshold of a node depends on its distance from initiators. 
To that end, the tree approximation used in this study can be applied.
This study also assumes that a fraction of initiators is small.
For the original Watts model, there exists a critical fraction of initiators at which a discontinuous transition from $\rho_\infty \approx \rho_0$ to $\rho_\infty \approx 1$ occurs~\cite{singh2013threshold,karampourniotis2015impact} as $\rho_0$ increases.
Cascade dynamics with a large fraction of initiators shows complex behaviors in the extended Watts model, which will be reported in the future.

\section*{Acknowledgement}
We thank Shogo Mizutaka, Tomokatsu Onaga, and Naoya Fujiwara for their useful comments and suggestions. This work was supported by JSPS KAKENHI Grant Numbers JP18KT0059, JP19K03648, and JP21H03425.

\appendix
\section{Derivation of Eqs.~(\ref{eq:rnEq}) and (\ref{Cn})}

In this Appendix, we derive Eq.~(\ref{eq:rnEq}) and Eq.~(\ref{Cn}) from Eq.~(\ref{eq:recEq-rl-alt}).
First, we rewrite Eq.~(\ref{eq:recEq-rl-alt}) as
\begin{eqnarray}
r_{l+1}=G_1(r_l)+G_2(r_l), \label{eq:start}
\end{eqnarray}
where
\begin{eqnarray}
G_1(r_l)= \sum_{k=0}^{\infty} \frac{(k+1)p(k+1)}{\langle k\rangle} \sum_{m=0}^k \binom{k}{m} [(1-\rho_0)(1-r_l)]^{k-m} [\rho_0+(1-\rho_0)r_l]^m F_\ea \left(\frac{m}{k+1}\right) \label{G1}
\end{eqnarray}
and
\begin{eqnarray}
G_2(r_l)=\sum_{k=0}^{\infty} \frac{(k+1)p(k+1)}{\langle k\rangle}\left(1-\rho_{0}\right)^{k} \sum_{m=0}^{k} \binom{k}{m} r_l^{m}(1-r_l)^{k-m} \left[F_\ma\left(\frac{m}{k+1}\right) - F_\ea\left(\frac{m}{k+1}\right)\right]. \label{G2}
\end{eqnarray}
We perform a second-order Taylor expansion of $G_1(r_l)$ at $r_l\simeq0$:
\begin{eqnarray}
G_1(r_l)
&\simeq&\sum_{k=0}^{\infty} \frac{(k+1)p(k+1)}{\langle k\rangle} \sum_{m=0}^{k} \binom{k}{m}\, \rho_0^{m} \left(1-\rho_{0}\right)^{k-m} F_\ea\left(\frac{m}{k+1}\right) 
\nonumber  \\
&&+\sum_{k=1}^{\infty} \frac{(k+1)p(k+1)}{\langle k\rangle} \left[-\sum_{m=0}^{k-1}\binom{k}{m}\, (k-m) \rho_0^{m} \left(1-\rho_{0}\right)^{k-m}  F_\ea\left(\frac{m}{k+1}\right)\right.
\nonumber \\
&& +\left. \sum_{m=1}^{k} \binom{k}{m}\, m \rho_0^{m-1} \left(1-\rho_{0}\right)^{k+1-m}  F_\ea\left(\frac{m}{k+1}\right)\right] r_l
\nonumber \\
&&+\sum_{k=2}^{\infty} \frac{(k+1p(k+1)}{\langle k\rangle} \left[ \frac{1}{2}\sum_{m=0}^{k-2} \binom{k}{m}\,  (k-m)(k-1-m)\rho_0^{m} \left(1-\rho_{0}\right)^{k-m}F_\ea\left(\frac{m}{k+1}\right)\right.
 \nonumber \\
&& -\sum_{m=1}^{k-1}\binom{k}{m}\, (k-m)m\rho_0^{m-1} \left(1-\rho_{0}\right)^{k+1-m}F_\ea\left(\frac{m}{k+1}\right) 
\nonumber \\
&&+\left.\frac{1}{2} \sum_{m=2}^{k} \binom{k}{m}\, m(m-1) \rho_0^{m-2}  \left(1-\rho_{0}\right)^{k+2-m}F_\ea\left(\frac{m}{k+1}\right)\right] r_l^2\nonumber \\
&=&\sum_{n=0}^2\sum_{k=n}^{\infty} \frac{(k+1)p(k+1)}{\langle k\rangle}\sum_{i=0}^n\sum_{m=0}^{k-n} \binom{k}{m+i} \binom{k-m-i}{n-i}\binom{m+i}{i}(-1)^{i+n}\rho_0^{m} \left(1-\rho_{0}\right)^{k-m}F_\ea\left(\frac{m+i}{k+1}\right)r_l^n \nonumber \\
&=&\sum_{n=0}^2\sum_{k=n}^{\infty} \frac{(k+1)p(k+1)}{\langle k\rangle}\sum_{i=0}^n\sum_{m=0}^{k-n}\binom{k}{n} \binom{k-n}{m}\binom{n}{i}(-1)^{i+n}\rho_0^{m} \left(1-\rho_{0}\right)^{k-m}F_\ea\left(\frac{m+i}{k+1}\right)r_l^n . \label{G1a}
\end{eqnarray}
A second-order Taylor expansion is also performed for $G_2(r_l)$ at $r_l\simeq0$:
\begin{equation}
G_2(r_l)\simeq
\sum_{n=0}^2\sum_{k=0}^{\infty} \frac{(k+1)p(k+1)}{\langle k\rangle}\sum_{i=0}^n\binom{k}{n}\binom{n}{i}(-1)^{i+n}\left(1-\rho_0\right)^{k} \left[F_\ma \left(\frac{i}{k+1}\right)-F_\ea\left(\frac{i}{k+1}\right) \right] r_l^n . \label{G2a}
\end{equation}
Substituting Eqs.~(\ref{G1a}) and (\ref{G2a}) into Eq.~(\ref{eq:start}), we have
\begin{eqnarray}
r_{l+1}&\approx&\sum_{n=0}^2\sum_{k=n}^{\infty} \frac{(k+1)p(k+1)}{\langle k\rangle}\sum_{i=0}^n\sum_{m=0}^{k-n}\binom{k}{n} \binom{k-n}{m}\binom{n}{i}(-1)^{i+n}\rho_0^{m} \left(1-\rho_{0}\right)^{k-m}F_\ea\left(\frac{m+i}{k+1}\right)r_l^n  \nonumber \\
& &+\sum_{n=0}^2\sum_{k=0}^{\infty} \frac{(k+1)p(k+1)}{\langle k\rangle}\sum_{i=0}^n\binom{k}{n}\binom{n}{i}(-1)^{i+n}\left(1-\rho_0\right)^{k} \left[F_\ma \left(\frac{i}{k+1}\right)-F_\ea\left(\frac{i}{k+1}\right) \right] r_l^n \nonumber
\\
&=&\sum_{n=0}^2\sum_{k=n}^{\infty} \frac{(k+1)p(k+1)}{\langle k\rangle}\sum_{i=0}^n\sum_{m=1}^{k-n}\binom{k}{n} \binom{k-n}{m}\binom{n}{i}(-1)^{i+n}\rho_0^{m} \left(1-\rho_{0}\right)^{k-m}F_\ea\left(\frac{m+i}{k+1}\right)r_l^n  \nonumber \\
& &+\sum_{n=0}^2\sum_{k=0}^{\infty} \frac{(k+1)p(k+1)}{\langle k\rangle}\sum_{i=0}^n\binom{k}{n}\binom{n}{i}(-1)^{i+n}\left(1-\rho_0\right)^{k}F_\ma \left(\frac{i}{k+1}\right)r_l^n \label{G1+G2}
\nonumber
\\
&=&\sum_{n=0}^2 C_n r_l^n, \label{eq:ap}
\end{eqnarray}
where
\begin{eqnarray}
C_n&=&
\sum_{k=n}^{\infty} \frac{(k+1)p(k+1)}{\langle k\rangle}\sum_{i=0}^n\sum_{m=1}^{k-n}\binom{k}{n} \binom{k-n}{m}\binom{n}{i}(-1)^{i+n}\rho_0^{m} \left(1-\rho_{0}\right)^{k-m}F_\ea\left(\frac{m+i}{k+1}\right)  \nonumber \\
&&+\sum_{k=0}^{\infty} \frac{(k+1)p(k+1)}{\langle k\rangle}\sum_{i=0}^n\binom{k}{n}\binom{n}{i}(-1)^{i+n}\left(1-\rho_0\right)^{k}F_\ma \left(\frac{i}{k+1}\right).
\end{eqnarray}
The last equations correspond with Eqs.~(\ref{eq:rnEq}) and (\ref{Cn}).




\end{document}